\documentclass[epj,final]{svjour}

\usepackage{latexsym}
\usepackage{url}
\usepackage{amsfonts}
\usepackage{longtable}

\RequirePackage{graphicx}

\begin{document}

\title{On the Earth's tidal perturbations for the LARES satellite}

\author{V.G. Gurzadyan\inst{1} \and I. Ciufolini\inst{2,3} \and H.G. Khachatryan\inst{1} \and S. Mirzoyan\inst{1} \and A. Paolozzi\inst{3,4}
\and G. Sindoni\inst{3,4}
}
\institute{Center for Cosmology and Astrophysics, Alikhanian National Laboratory and Yerevan State University, Yerevan, Armenia  \and Dipartimento di Ingegneria dell Innovazione, Universita del Salento, Lecce, Italy \and Centro Fermi, Rome, Italy \and SIA, Sapienza Universita di Roma, Rome, Italy 
}
 
\date{Received: date / Revised version: date}

\abstract{
Frame dragging, one of the outstanding phenomena predicted by General Relativity, is efficiently studied by means of the laser-ranged satellites LARES, LAGEOS and LAGEOS 2. The accurate analysis of the orbital perturbations of Earth's solid and ocean tides has been relevant for increasing the accuracy in the test of frame-dragging using these three satellites. The Earth's tidal perturbations acting on the LARES satellite are obtained for the 110 significant modes of corresponding Doodson number and are exhibited to enable the comparison to those of the LAGEOS and LAGEOS-2 satellites. For LARES we represent 29 perturbation modes for $l=2,3,4$ for ocean tides.} 

\PACS{
      {04.80.-y}{Experimental studies of gravity}
}     
\maketitle

\section*{Introduction}

Frame-dragging is one of the intriguing phenomena of General Relativity with relevant astrophysical implications \cite{CW} and its study and test with increased accuracy using laser-ranged satellites is a goal of profound importance \cite{Ciu2007}.  The LARES (LAser RElativity Satellite) satellite, as the best ever made test-particle to move on geodesics in the Earth's gravitational field \cite{Ciu2012},\cite{Ciu2013}, is proving its efficiency for high precision probing of General Relativity and fundamental physics \cite{Ciu2015}. The analysis of 3.5 years of laser ranging data of LARES, together with those of the two LAGEOS satellites, has resulted in a test of frame-dragging with about 5\% accuracy\cite{Ciu2016}. To extract frame-dragging from the laser-ranging data, the main gravitational and non-gravitational orbital perturbations are modeled in orbital estimators such as GEODYN, EPOSOC and UTOPIA. The analysis of the laser-ranging data using these orbital estimators produces the post-fit orbital residuals of the orbital elements, i.e. the difference between the observed orbital elements and the modeled ones. Then, a relevant part of the analysis of the post-fit residuals of LARES, LAGEOS and LAGEOS 2, and of the relative error analysis, was based on the study of the main orbital perturbations due to the Earth tides and to the non-gravitational effects, and in particular on the analysis of their periods and amplitudes on the nodes of LARES, LAGEOS and LAGEOS 2. The main tidal signals were then fitted for, together with frame-dragging, in the combination of the post-fit residuals of the nodes of LAGEOS, LAGEOS 2 and LARES \cite{Ciu2016}. 

The analysis of the orbital perturbations of a satellite due to the main tide modes of the Earth's can be performed by means of the perturbative methods developed in celestial mechanics \cite{Br}. Below, we compute the parameters of the 110 modes in Doodson number classification, representing them along with those of LAGEOS and LAGEOS-2 satellites in mutual comparison in relevance and statistics of the modes.

\section{Tidal perturbations}

\subsection{Earth tides}
To describe the satellite's motion influenced by the tides one must adopt a model for the gravitational
potential of the Earth. Earth's shape is not spherical, and although the
quadrupole has essential contribution in the perturbed potential, such effects as the tidal 
deformations of the oceans and of the atmosphere have to be properly accounted for. 

Below we briefly summarize the main steps in the analysis. Earth's gravitational potential is convenient to represent as expansion by spherical harmonics $Y_{lm}(\phi, \lambda)$ \cite{kaula,carp89,Io,iers2010} 
\begin{eqnarray}
V(r,\phi,\lambda)&=&\frac{GM}{R}\sum_{l=0}^{\infty}\left(\frac{R}{r}\right)^{l+1}\sum_{m=-l}^{l}\frac{w_{lm}Y_{lm}(\phi,\lambda)}{2-\delta_{0m}} \nonumber \\
w^{*}_{lm}&=&w_{l-m}=C_{lm} + iS_{lm}\delta_{0m}
\end{eqnarray}
here $R,M$ are Earth's mean equatorial radius and mass, respectively. Coefficients $C_{lm},S_{lm}$ are so-called Stokes coefficients.

The gravitational perturbations by the Moon and Sun are the essential ones in inducing Earth's tides. 
The tidal perturbation potential can be written as function of six orbital parameters $\{a,i,e,\omega,\mu,\Omega\}$ of the satellite that varying with time \cite{kaula} (Eq.(3.70) there), \cite{cata}
\begin{equation}
V_{lm}(a,i,e,\omega,\mu,\Omega)=\frac{GM}{R}\left(\frac{R}{a}\right)^{l+1}\sum_{p=0}^{l}F_{lmp}(i)\sum_{q=-\infty}^{\infty}G_{lpq}(e)S_{lmpq}(\omega,\mu,\Omega,\theta)
\end{equation}
where $a, F_{lmp}(i), G_{lpq}(e)$ are the semi-major axis, inclination and eccentricity functions of the perturbation, respectively. In general, both have complicated form but in the case of our interest,
$l=2,\,m=0,\,1,\,2$, they read  
\begin{eqnarray}
F_{201}(i) &=& \frac{3}{4}sin^{2}i - \frac{1}{2}  \nonumber \\
F_{211}(i) &=& -\frac{3}{2}sin\, i\,\, cos\, i  \nonumber \\
F_{221}(i) &=& -\frac{3}{2}sin^{2}i  \nonumber \\
G_{210}(e) &=& \frac{1}{\sqrt{(1-e^2)^3}}
\end{eqnarray}
The angular perturbation function is \cite{kaula,carp89,iers2010}
\begin{eqnarray}
N_{lm} &=& \sqrt{\frac{2l+1}{4\pi}\frac{(l-m)!}{(l+m)!}} \nonumber \\
S_{lmpq}(\omega,\mu,\Omega,\theta) &=& N_{lm}\sum_{i}{u_{lm}k_{lm}(\nu_{i})cos(\nu_{i}\,t+\xi_{lmpq})} \nonumber \\
\xi_{lmpq} &=& (l-2p)\omega+(l-2p+q)\mu+m(\Omega-\theta)+\eta_{lm}(\nu)
\label{slm}
\end{eqnarray}
here $\omega,\mu,\Omega,\theta$ are the argument of the perigee, mean anomaly, the longitude of ascending node and sidereal time, respectively. Normalization coefficient $N_{lm}$ arises from common notation of spherical harmonics.

Tidal potential harmonic coefficients $u_{lm}$ are listed and computed in \cite{cata},\cite{caed},\cite{doodson}. They are coefficients of time-harmonic expansion of tide potential. The Love numbers manifest the fact that tides can deform the Earth's shape. They measure response of the Earth to tidal potential. The Love numbers $k_{lm}(\nu)$ weakly depend on the frequency of tidal perturbation mode $\nu$ \cite{love},\cite{iers2010}. Love numbers for about 110 modes are listed in 
\cite{iers2010} for $l=0,\,1,\,2$. It has a general form 
\begin{eqnarray}
k_{l} &=& k_{l}^{real} + i\,k_{l}^{imag}  \nonumber \\
\delta k_{lm}(\nu) &=& \delta k_{lm}^{real}(\nu) + i\,\delta k_{lm}^{imag}(\nu) \nonumber \\
k_{lm}(\nu) &=& \left|k_{l} + \delta k_{lm}(\nu)\right|
\end{eqnarray}
And phase $\eta_{lm}(\nu)$ in eq. \ref{slm} has form
\begin{equation}
\eta_{lm}(\nu)=\arctan\left(\frac{k_{l}^{imag} + \delta k_{lm}^{imag}(\nu)}{k_{l}^{real} + \delta k_{lm}^{real}(\nu)} \right)
\end{equation}
Each mode is represented by Doodson's number  $\left\{i_1,\ldots,i_6\right\}$ \cite{doodson}. To get real values of Doodson numbers the vector of the numbers $\left\{0,\,5,\,5,\,5,\,5,\,5\right\}$ should be subtracted. Such a notation has been adopted to prevent negative values in Doodson number vector and for easy representation e.g. $055.565$ without minus $(-)$ sign. Each Doodson number is grouped by $(.)$ into high and low frequencies. For example, $055.565(000.010)$ implies high period ($period=6798\,days$) mode. In general, the period of perturbation depends on the nodal frequency of the satellite $\dot{\Omega}$ and periodic behavior of five parameters: 1) mean lunar time, 2) Lunar mean longitude, 3) Solar mean longitude, 4) longitude of Lunar perigee, 5) longitude of Lunar ascending node, 6) longitude of Solar perigee. The periods for each of these six parameters are given in \cite{doodson} as a dimensionless parameter $\alpha_{i}=\left\{347.80925061,\,13.17639673,\,0.98564734,\,0.11140408,\,0.05295392,\,0.00004707\right\}$ and are defined as 
\begin{equation}
\gamma_{i}=\frac{360}{\alpha_{i}}\,days.
\end{equation}

The frequency of the mode is \cite{carp89}

\begin{equation}
\dot{\Gamma}_{j,lmp}=\sum_{i=1}^{6}\frac{2{\pi}j_i}{\gamma_i}+m(\dot{\Omega}-\frac{2\pi}{\gamma_1})+(l-2p){\dot{\omega}}+(l-2p+q){\dot{\mu}},
\end{equation}
here $\dot{\Omega},\dot{\omega}$ and $\dot{\mu}$ represent the frequencies of the longitude of ascending node, argument of the perigee and mean anomaly for the satellite, respectively. 

From the Lagrange equation \cite{kaula} one has for the ascending node $\Omega$ 
\begin{equation}
\frac{d\Omega}{dt}=\frac{1}{\sqrt{GMa(1-e^2)}}\frac{\partial{V}}{\partial{i}}.
\label{omegadot}
\end{equation}

Since the derivative of the potential $V$ is taken over the inclination parameter $i$ and only the argument of {\it cos} in $V$ depends on time, the main perturbation of the longitude of the ascending node arises due to the variation of {\it cos}. In view of this, after formal integration in linear approximation one gets for the amplitude of perturbations (cf. Eq. (3.76) in \cite{kaula} and Eq. (3.8) in \cite{carp89})

\begin{eqnarray}
\Delta \Omega_{lmpq}  &=&\frac{180}{\pi{\sin(i)}}\sqrt{\frac{GMR^{2l-2}}{a^{2l+3}(1-e^{2})}\frac{2l+1}{4\pi}\frac{(l-m)!}{(l+m)!}}\frac{dF_{lmp}(i)}{di}G_{lpq}(e)\frac{u_{lm}k_{lm}(\nu )}{\dot{\Gamma}_{j,lmp}}.
\end{eqnarray}
 
We are interested in the modes with long period without mean anomaly of satellite $\mu$. Because of that we set $l=2,p=1,q=0,m=0,1,2$. Some other possible combinations of parameters are omited because of equation $G_{20-2}(e)=G_{222}(e)=0$. In addition, one has the following parameters of the orbit of the LARES satellite: 

\begin{eqnarray}
a_{L} &=& 7820\,km  \nonumber \\
e_{L} &=& 0.0008  \nonumber \\
i_{L} &=& 69.5^{o}  \nonumber \\
P_{L} &=& 115\, min  \nonumber \\
P({\Omega}_{L}) &=& -211\,days \nonumber \\
P({\omega}_{L}) &=& -382\,days
\end{eqnarray}

The results of the computations for 110 significant tides for these parameters of LARES's orbit are given in the Table 1 (cf. \cite{ijmpd}).

\begin{longtable}[B]{lrrrrr}
\multicolumn{6}{c}{{\bf Table 1}. Amplitudes $\Delta\Omega$ and periods of perturbations for the LARES satellite}\\
\multicolumn{6}{c}{generated by Moon and Sun induced tides of the Earth.}\\
\multicolumn{6}{c}{$l=2$,$m=0$,$p=1$,$q=0$}\\
\hline
 & Mode & Love number & Period(days) & $u_{lm}$ & $\Delta\Omega(mas)$ \\
\hline\endfirsthead
 & Mode & Love number & Period(days) & $u_{lm}$ & $\Delta\Omega(mas)$ \\
\hline\endhead
\hline
\multicolumn{6}{c}{\textit{Continued on next page}}
\endfoot
\hline\endlastfoot
            & 055.565 & 0.315416 & 6798.3636 &  0.02793 & 5359.6967\\
            & 055.575 & 0.313178 & 3399.1818 & -0.00027 & -25.7223 \\
 $S_a$      & 056.554 & 0.307390 &  365.2596 & -0.00492 & -49.4353 \\
 $S_{sa}$   & 057.555 & 0.305946 &  182.6211 & -0.031   & -155.0024\\
            & 057.565 & 0.305896 &  177.8438 & 0.00077  & 3.7487   \\
            & 058.554 & 0.305174 &  121.7493 & -0.00181 & -6.0183  \\
 $M_{sm}$   & 063.655 & 0.302920 &   31.8119 & -0.00673 & -5.8038  \\
            & 065.445 & 0.302709 &   27.6667 & 0.00231  & 1.7313   \\
 $M_m$      & 065.455 & 0.302709 &   27.5546 & -0.03518 & -26.2600 \\
            & 065.465 & 0.302699 &   27.4433 & 0.00229  & 1.7024   \\
            & 065.655 & 0.302679 &   27.0925 & 0.00188  & 1.3797   \\
 $M_{sf}$   & 073.555 & 0.301818 &   14.7653 & -0.00583 & -2.3251  \\
            & 075.355 & 0.301728 &   13.7773 & -0.00288 & -1.0714  \\
 $M_f$      & 075.555 & 0.301718 &   13.6608 & -0.06663 & -24.5769 \\
            & 075.565 & 0.301718 &   13.6334 & -0.02762 & -10.1674 \\
            & 075.575 & 0.301718 &   13.6061 & -0.00258 & -0.9478  \\
 $M_{stm}$  & 083.655 & 0.301257 &    9.5569 & -0.00242 & -0.6235  \\
 $M_{tm}$   & 085.455 & 0.301197 &    9.1329 & -0.01276 & -3.1412  \\
            & 085.465 & 0.301197 &    9.1207 & -0.00529 & -1.3005  \\
 $M_{sqm}$  & 093.555 & 0.300886 &    7.0958 & -0.00204 & -0.3898  \\
 $M_{qm}$   & 095.355 & 0.300846 &    6.8594 & -0.00169 & -0.3121  \\
\end{longtable}

\begin{longtable}{lrrrrr}
\multicolumn{6}{c}{{\bf Table 2}. Amplitudes $\Delta\Omega$ and periods of perturbations for the LARES satellite}\\
\multicolumn{6}{c}{generated by Moon and Sun induced tides of the Earth.}\\
\multicolumn{6}{c}{$l=2$,$m=1$,$p=1$,$q=0$}\\
\hline
 & Mode & Love number & Period(days) & $u_{lm}$ & $\Delta\Omega(mas)$ \\
\hline\endfirsthead
 & Mode & Love number & Period(days) & $u_{lm}$ & $\Delta\Omega(mas)$ \\
\hline\endhead
\hline
\multicolumn{6}{c}{\textit{Continued on next page}}
\endfoot
\hline\endlastfoot
   ${2Q}_1$  & 125.755 & 0.298013 & -6.6434   & -0.006640 & 1.1050 \\
 $\sigma_1$ & 127.555 & 0.298003 & -6.8649   & -0.008020 & 1.3791 \\
            & 135.645 & 0.297853 & -8.7428   & -0.009470 & 2.0729 \\
 $Q_1$      & 135.655 & 0.297843 & -8.7540   & -0.050200 & 11.0020 \\
 $\rho_1$   & 137.455 & 0.297813 & -9.1428   & -0.009540 & 2.1834 \\
            & 145.545 & 0.297483 & -12.8060  & -0.049460 & 15.8380 \\
 $O_1$      & 145.555 & 0.297473 & -12.8301  & -0.262210 & 84.1201 \\
 $\tau_1$   & 147.555 & 0.297393 & -13.7996  & 0.003430  & -1.1832 \\
${N\tau}_1$ & 153.655 & 0.296623 & -21.5022  & 0.001940  & -1.0401 \\
            & 155.445 & 0.296373 & -23.9252  & 0.001370  & -0.8166 \\
 ${Lk}_1$   & 155.455 & 0.296363 & -24.0097  & 0.007410  & -4.4320 \\
 ${No}_1$   & 155.655 & 0.296333 & -24.3718  & 0.020620  & -12.5178 \\
            & 155.665 & 0.296323 & -24.4595  & 0.004140  & -2.5222 \\
 $\chi_1$   & 157.455 & 0.295993 & -27.6441  & 0.003940  & -2.7099 \\
            & 157.465 & 0.295973 & -27.7570  & 0.000870  & -0.6008 \\
 $\pi_1$    & 162.556 & 0.289961 & -77.2026  & -0.007140 & 13.4351 \\
            & 163.545 & 0.287131 & -96.5041  & 0.001370  & -3.1909 \\
 $P_1$      & 163.555 & 0.286921 & -97.8938  & -0.122030 & 288.1082 \\
            & 164.554 & 0.280660 & -133.7367 & 0.001030  & -3.2497 \\
 $S_1$      & 164.556 & 0.280660 & -133.7414 & 0.002890  & -9.1184 \\
            & 165.345 & 0.267821 & -181.6417 & 0.000070  & -0.2862 \\
            & 165.535 & 0.262000 & -198.6679 & 0.000050  & -0.2188 \\
            & 165.545 & 0.259851 & -204.6484 & -0.007300 & 32.6308 \\
 $K_1$      & 165.555 & 0.257463 & -211.0000 & 0.368780  & -1683.9767 \\
            & 165.565 & 0.254755 & -217.7585 & 0.050010  & -233.1986 \\
            & 165.575 & 0.251757 & -224.9644 & -0.001080 & 5.1415 \\
            & 166.455 & 1.176585 & -432.7350 & -0.000004 & 0.1592 \\
            & 166.544 & 0.650619 & -465.4080 & 0.000009  & -0.2312 \\
 $\psi_1$   & 166.554 & 0.526244 & -499.6108 & 0.002930  & -64.7528 \\
            & 166.556 & 0.526104 & -499.6761 & -0.000042 & 0.9305 \\
            & 166.564 & 0.466726 & -539.2394 & 0.000050  & -1.0578 \\
            & 167.355 & 0.335867 & 8505.5345 & 0.000180  & 43.2228 \\
            & 167.365 & 0.333836 & 3778.3652 & 0.000060  & 6.3615 \\
 $\phi_1$   & 167.555 & 0.328564 & 1357.8063 & 0.005250  & 196.8743 \\
            & 167.565 & 0.327234 & 1131.7642 & -0.000200 & -6.2261 \\
            & 168.554 & 0.314689 & 287.8310  & 0.000310  & 2.3602 \\
 $\theta_1$ & 173.655 & 0.302004 & 37.4596   & 0.003950  & 3.7562 \\
            & 173.665 & 0.301994 & 37.2544   & 0.000780  & 0.7376 \\
            & 175.445 & 0.301554 & 31.8418   & -0.000600 & -0.4843 \\
 $J_1$      & 175.455 & 0.301544 & 31.6934   & 0.020620  & 16.5646 \\
            & 175.465 & 0.301534 & 31.5463   & 0.004090  & 3.2702 \\
 ${So}_1$   & 183.555 & 0.300244 & 15.8763   & 0.003420  & 1.3703 \\
            & 185.355 & 0.300154 & 14.7397   & 0.001690  & 0.6285 \\
 ${Oo}_1$   & 185.555 & 0.300144 & 14.6065   & 0.011290  & 4.1604 \\
            & 185.565 & 0.300144 & 14.5751   & 0.007230  & 2.6586 \\
            & 185.575 & 0.300144 & 14.5440   & 0.001510  & 0.5541 \\
 $\nu_1$    & 195.455 & 0.299714 & 9.5461    & 0.002160  & 0.5195 \\
            & 195.465 & 0.299714 & 9.5327    & 0.001380  & 0.3314 \\
\end{longtable}

\newpage

\begin{longtable}{lrrrrr}
\multicolumn{6}{c}{{\bf Table 3}. Amplitudes $\Delta\Omega$ and periods of perturbations for the LARES satellite}\\
\multicolumn{6}{c}{generated by Moon and Sun induced tides of the Earth.}\\
\multicolumn{6}{c}{$l=2$,$m=2$,$p=1$,$q=0$}\\
\hline
 & Mode & Love number & Period(days) & $u_{lm}$ & $\Delta\Omega(mas)$ \\
\hline\endfirsthead
 & Mode & Love number & Period(days) & $u_{lm}$ & $\Delta\Omega(mas)$ \\
\hline\endhead
\hline
\multicolumn{6}{c}{\textit{Continued on next page}}
\endfoot
\hline\endlastfoot
            & 225.855 & 0.301083 & -5.2204   & 0.001800  & -0.1034 \\
            & 227.655 & 0.301083 & -5.3562   & 0.004670  & -0.2751 \\
            & 235.755 & 0.301083 & -6.4406   & 0.016010  & -1.1343 \\
            & 237.555 & 0.301083 & -6.6486   & 0.019320  & -1.4130 \\
            & 245.555 & 0.301083 & -8.3835   & -0.003890 & 0.3587 \\
            & 245.645 & 0.301083 & -8.3949   & -0.004510 & 0.4165 \\
 $N_2$      & 245.655 & 0.301083 & -8.4053   & 0.120990  & -11.1865 \\
            & 247.455 & 0.301063 & -8.7630   & 0.022980  & -2.2150 \\
            & 253.755 & 0.301063 & -11.4236  & -0.001900 & 0.2387 \\
            & 254.556 & 0.301063 & -11.7070  & -0.002180 & 0.2807 \\
            & 255.545 & 0.301063 & -12.0732  & -0.023580 & 3.1313 \\
 $M_2$      & 255.555 & 0.301063 & -12.0947  & 0.631920  & -84.0658 \\
            & 256.554 & 0.301063 & -12.5089  & 0.001920  & -0.2642 \\
            & 263.655 & 0.301063 & -19.5137  & -0.004660 & 1.0002 \\
            & 265.455 & 0.301063 & -21.5567  & -0.017860 & 4.2347 \\
            & 265.555 & 0.301063 & -21.7015  & 0.003590  & -0.8569 \\
            & 265.655 & 0.301063 & -21.8482  & 0.004470  & -1.0742 \\
            & 265.665 & 0.301063 & -21.9187  & 0.001970  & -0.4749 \\
            & 271.557 & 0.301063 & -48.9475  & 0.000700  & -0.3769 \\
 $T_2$      & 272.556 & 0.301063 & -56.5219  & 0.017200  & -10.6932 \\
 $S_2$      & 273.555 & 0.301063 & -66.8695  & 0.294000  & -216.2409 \\
            & 273.755 & 0.301063 & -69.7565  & 0.000040  & -0.0307 \\
            & 274.554 & 0.301063 & -81.8551  & -0.002460 & 2.2148 \\
            & 274.556 & 0.301063 & -81.8568  & 0.000620  & -0.5582 \\
            & 274.566 & 0.301063 & -82.8545  & -0.000040 & 0.0365 \\
            & 273.655 & 0.301063 & -68.2825  & 0.000040  & -0.0300 \\
            & 275.455 & 0.301063 & -102.1646 & 0.000190  & -0.2135 \\
            & 275.465 & 0.301063 & -103.7233 & 0.000040  & -0.0456 \\
            & 275.545 & 0.301063 & -103.8878 & 0.001030  & -1.1770 \\
 $K_2$      & 275.555 & 0.301063 & -105.5000 & 0.079960  & -92.7870 \\
            & 275.565 & 0.301063 & -107.1630 & 0.023830  & -28.0887 \\
            & 277.555 & 0.301063 & -249.8217 & 0.000630  & -1.7311 \\
            & 275.575 & 0.301063 & -108.8793 & 0.002590  & -3.1018 \\
            & 282.656 & 0.301063 & 52.0338   & 0.000040  & 0.0229 \\
            & 283.445 & 0.301063 & 47.1919   & 0.000060  & 0.0311 \\
            & 283.455 & 0.301063 & 46.8666   & 0.000040  & 0.0206 \\
            & 283.655 & 0.301063 & 45.5455   & 0.000850  & 0.4258 \\
            & 283.665 & 0.301063 & 45.2424   & 0.000370  & 0.1841 \\
            & 283.675 & 0.301063 & 44.9433   & 0.000040  & 0.0198 \\
            & 285.455 & 0.301063 & 37.2954   & 0.004470  & 1.8337 \\
            & 285.465 & 0.301063 & 37.0919   & 0.001950  & 0.7956 \\\hline
\end{longtable}

Figures 1, 2 represent the frequency counts for the amplitude, period for the 110 tide modes and Figure 3 shows the amplitude vs period;
the pink line denotes the frequency of the node of LARES satellite.

\begin{figure}[!htbp]
  \centering
  \includegraphics[width=100mm]{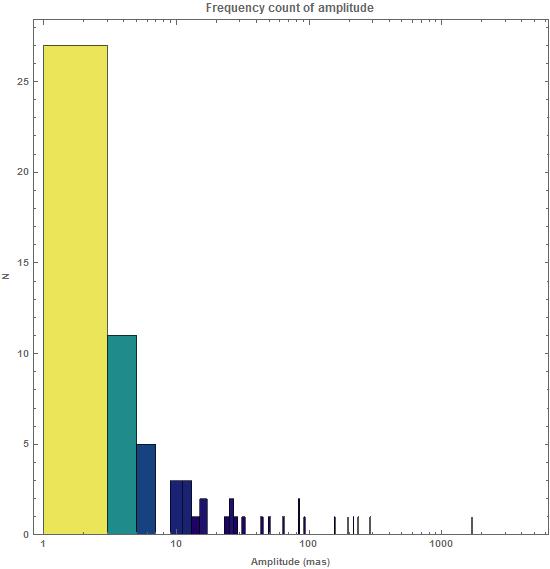}
  \label{fig:1}
  \caption{The frequency count for the amplitudes (in milliarcseconds: {\it mas}) of 110 significant tide modes of the LARES satellite.}
\end{figure}

\begin{figure}[!htbp]
  \centering
  \includegraphics[width=100mm]{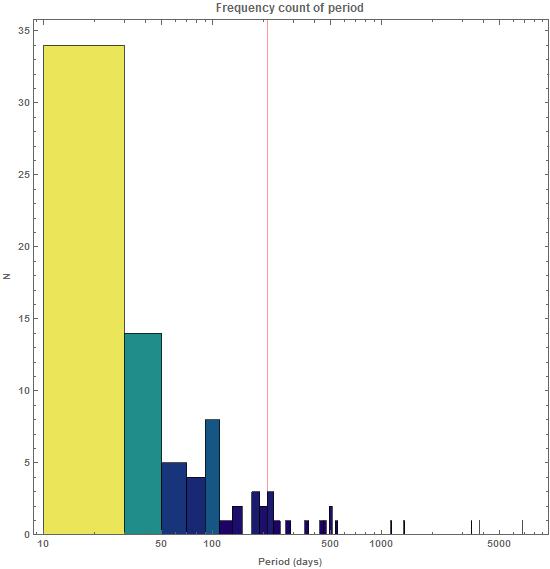}
  \label{fig:2}
  \caption{The frequency count for the periods (in days) of 110 tide modes for the LARES satellite.}
\end{figure}

\begin{figure}[!htbp]
  \centering
  \includegraphics[width=100mm]{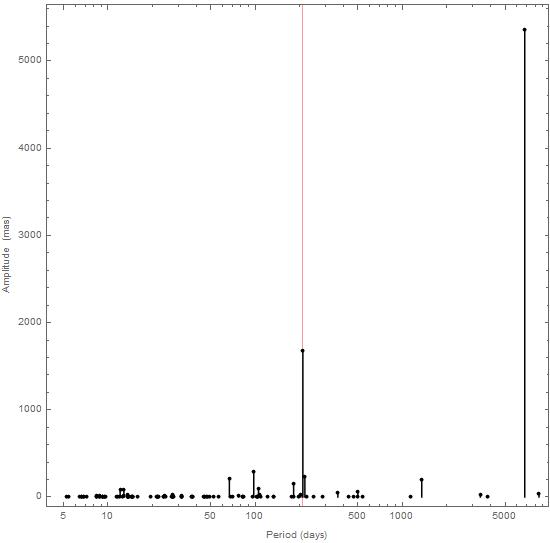}
  \label{fig:3}
  \caption{The amplitude vs period for the modes of Figure 1 and 2.}
\end{figure}

Similar estimations for the tides' amplitude, period and mutual dependence of the two latter parameters but for LAGEOS and LAGEOS-2, are exhibited in Figures 4-6 and 7-9, respectively. The following input parameters have been used for the orbits of LAGEOS and LAGEOS-2 satellites:

\begin{center}
\begin{tabular}{ccc}
Parameters     & LAGEOS        & LAGEOS-2       \\ \hline
{\it a}              & 12270\,km       & 12163\,km      \\
{\it e}              & 0.0045          & 0.0135         \\
{\it i}              & $109.84^{o}$    &  $52.64^{o}$   \\
{\it P}              & 0.1566 days  & 0.1545 days \\
${\it P(\Omega)}$ & 1,043.67 days & -569.21 days \\
${\it P(\omega)}$ & 1,707.62 days & 821.79 days
\end{tabular}
\end{center}

\begin{figure}[!htbp]
  \centering
  \includegraphics[width=100mm]{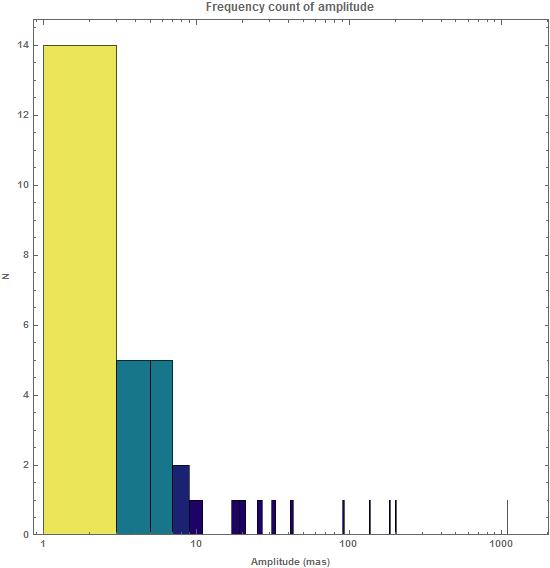}
  \label{fig:4}
  \caption{The same as in Figure 1, but for LAGEOS satellite.}
\end{figure}

\begin{figure}[!htbp]
  \centering
  \includegraphics[width=100mm]{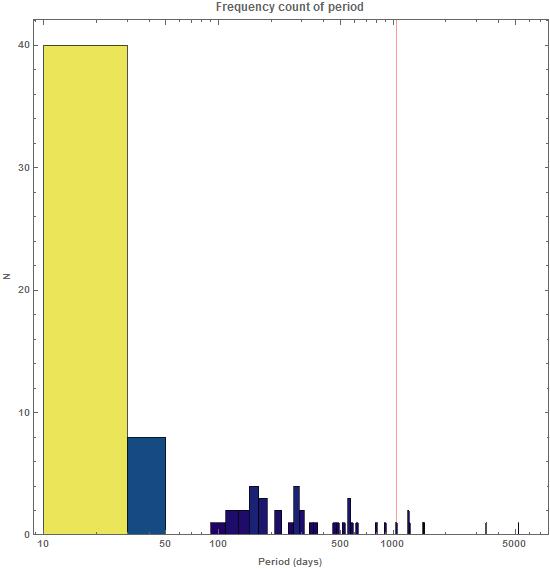}
  \label{fig:5}
  \caption{The same as in Figure 2, but for LAGEOS satellite.}
\end{figure}

\begin{figure}[!htbp]
  \centering
  \includegraphics[width=100mm]{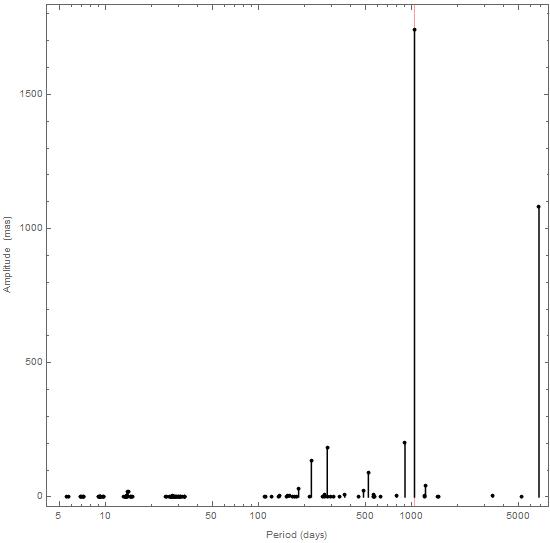}
  \label{fig:6}
  \caption{The same as in Figure 3, but for LAGEOS satellite.}
\end{figure}

\begin{figure}[!htbp]
  \centering
  \includegraphics[width=100mm]{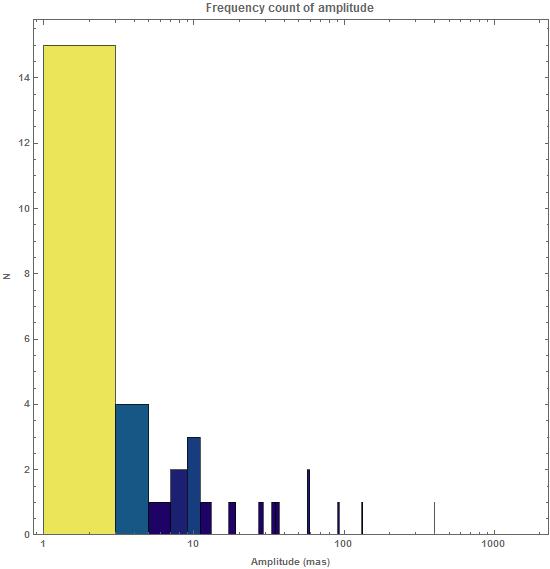}
  \label{fig:7}
  \caption{The same as in Figure 1, but for LAGEOS-2 satellite.}
\end{figure}

\begin{figure}[!htbp]
  \centering
  \includegraphics[width=100mm]{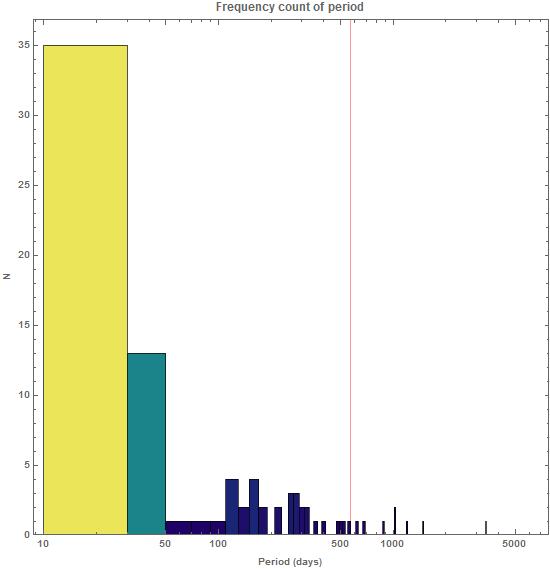}
  \label{fig:8}
  \caption{The same as in Figure 2, but for LAGEOS-2 satellite.}
\end{figure}

\begin{figure}[!htbp]
  \centering
  \includegraphics[width=100mm]{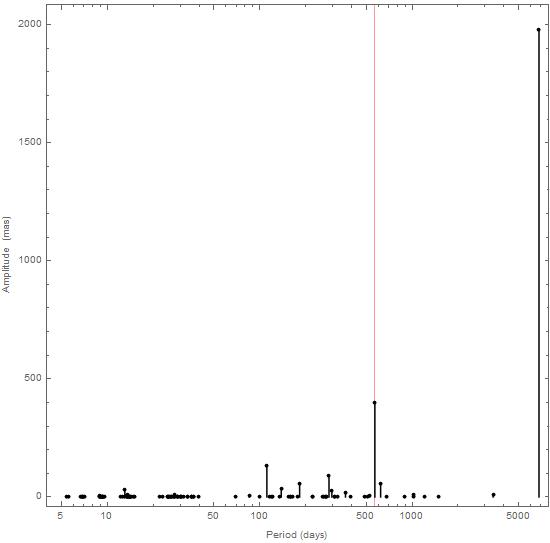}
  \label{fig:9}
  \caption{The same as in Figure 3, but for LAGEOS-2 satellite.}
\end{figure}

\subsection{Ocean tides}

The gravity of the Moon and Sun generates also ocean tides. The ocean tides can be described by the amplitude $\xi_a(\lambda,\phi)$ and phase $\psi(\lambda,\phi)$, both as a function of geocentric longitude and latitude ($\lambda$,$\phi$). 
\begin{equation}
\zeta(\lambda,\phi)=\xi_a(\lambda,\phi)cos(\theta_a + \chi - \psi(\lambda,\phi))
\label{height}
\end{equation}
Eq.(\ref{height}) can be represented via real valued series of prograde and retrograde waves of tides \cite{NSWC,Cart91,Ray1999,Wu2012}
\begin{equation}
\zeta(\phi,\lambda,t)=\sum_{l,m}\{D^{+}_{lm}\cos(\nu t + m \phi - \psi^{+}_{lm}) + D^{-}_{lm}\sin(\nu t - m \phi - \psi^{-}_{lm})\}P^{m}_{l}(\cos\lambda)
\end{equation}
Those waves can be treated as disturbance to the Earth's gravitational potential by the term \cite{NSWC,GEMT1,EGM96,FES2004}
\begin{equation}
V_{O}(r,\phi, \lambda)=4{\pi}G\rho_{w}R\sum_{l=0}^{\infty}\sum_{m=-l}^{l}\sum_{-}^{+}{\left(\frac{R}{r}\right)}^{l+1}\frac{(1+\tilde{k_l})D_{lm}^{\pm}}{2l+1}P_{lm}(\cos\lambda)\cos(\eta_{lm}^{\pm}(\nu))
\end{equation}
where $\rho_w=1023\,\,kg/m^3$ is the ocean water mean density.

With similar derivations as for the Earth's tides, one can define the perturbations of the ascending node induced by the ocean tides with amplitudes
\begin{equation}
{\Delta\Omega}^{O}_{lmpq}=\frac{720{\rho}_{w}}{\sin(i)}\sqrt{\frac{GR^{2l+4}}{Ma^{2l+3}(1-e^{2})}}\frac{dF_{lmp}(i)}{di}G_{lpq}(e)\frac{D_{lm}^{\pm}(1+\tilde{k}_l(\nu))}{(2l+1)\dot{\Gamma}_{j,lmp}}
\end{equation}
where $\tilde{k}_l(\nu)$ is the load Love number listed in \cite{Farrel72,NA11}. We list some values of $\tilde{k}_l(\nu)$

\begin{center}
\begin{tabular}{cccccc}
\hline
$\tilde{k}_l(\nu)$ & $l=1$   & $l=2$   & $l=3$    & $l=4$   & $l=5$ \\
                   & -0.3075 & -0.1950 & -0.1320  & -0.1032 & -0.0892\\
\hline
\end{tabular}
\end{center}

Several models for the ocean tides have been developed i.e. NSWC, EGM96, GOT99, FES2004 (see \cite{NSWC,EGM96,Ray1999,FES2004}). Among the two recent models - GOT99 and GOT4.7 - the latter uses more observational data and, particularly, it describes more accurately longer period ocean tides like the mode 056.554. Below we use GOT4.7 to obtain the ocean tides for the the LARES satellite. We are only interested in long periodic modes that do not contain mean anomaly $\mu$, and consider only the modes obeying the condition $l-2p+q=0$.

\begin{longtable}{lrrr}
\multicolumn{4}{c}{{\bf Table 4}. Amplitudes $\Delta\Omega$ and periods of the perturbations for the LARES satellite}\\
\multicolumn{4}{c}{by the Moon and Sun induced tides of the ocean.}\\
\multicolumn{4}{c}{$l=2$,$p=1$,$q=0$}\\
\hline
 & Mode & Period(days) & ${\Delta\Omega}_{O}(mas)$ \\
\hline\endfirsthead
 & Mode & Period(days) & ${\Delta\Omega}_{O}(mas)$ \\
\hline\endhead
\hline
\multicolumn{4}{c}{\textit{Continued on next page}}
\endfoot
\hline\endlastfoot
 $\Omega_1$    & 055.565 & 6798.3636 & 315.2685 \\
 $S_a$         & 056.554 & 365.2596  & 2.9845 \\
 $S_{sa}$      & 057.555 & 182.6211  & 9.3936 \\
 $M_m$         & 065.455 & 27.5546   & 1.4703 \\
 $M_f$         & 075.555 & 13.6608   & 1.2401 \\
 $M_{tm}$      & 085.455 & 9.1329    & 0.1592 \\
 $2Q_1$        & 125.755 & -6.6434   & -0.1130 \\
 $\sigma_1$    & 127.555 & -6.8649   & -0.1389 \\
 $Q_1$         & 135.655 & -8.7540   & -0.9950 \\
 $O_1$         & 145.555 & -12.8301  & -6.6186 \\
 $M_1$         & 155.655 & -24.3718  & -0.8842 \\
 $\pi_1$       & 162.556 & -77.2026  & -0.8812 \\
 $P_1$         & 163.555 & -97.8938  & -18.9394 \\
 $S_1$         & 164.555 & -133.7391 & -1.4276 \\
 $K_1$         & 165.555 & -211.0000 & -121.0278 \\
 $\psi_1$      & 166.554 & -499.6108 & -2.2555 \\
 $\phi_1$      & 167.555 & 1357.8063 & 10.8730 \\
 $J_1$         & 175.455 & 31.6934   & 0.8884 \\
 ${Oo}_1$      & 185.555 & 14.6065   & 0.1928 \\
 $2N_2$        & 235.755 & -6.4406   & -0.2312 \\
 $\mu_2$       & 237.555 & -6.6486   & -0.2831 \\
 $N_2$         & 245.655 & -8.4053   & -2.0030 \\
 $\nu_2$       & 247.455 & -8.7630   & -0.3898 \\
 $M_2$         & 255.555 & -12.0947  & -13.2661 \\
 $L_2$         & 265.455 & -21.5567  & -0.5843 \\
 $T_2$         & 272.556 & -56.5219  & -1.3064 \\
 $S_2$         & 273.555 & -66.8695  & -26.1269 \\
 $K_2$         & 275.555 & -105.5000 & -11.1390 \\
 $\eta_2$      & 285.455 & 37.2954   & 0.1798 \\
\end{longtable}
For modes with parameters ($l=3,p=1,q=-1$) the Eccentricity Function $G_{lpq}(e) \approx e$ for the eccentricity of LARES's orbit $e=0.0008$, therefore although $D_{lm}^{\pm}$ harmonic parameters of tides can be as high as for ($l=2,p=1,q=0$), their amplitudes are far smaller. 
\begin{longtable}{lrrr}
\multicolumn{4}{c}{{\bf Table 5}. Amplitudes $\Delta\Omega$ and periods of the perturbations for the LARES satellite}\\
\multicolumn{4}{c}{by the Moon and Sun induced tides of the ocean.}\\
\multicolumn{4}{c}{$l=3$,$p=1$,$q=-1$}\\
\hline
 & Mode & Period(days) & ${\Delta\Omega}_{O}(mas)$ \\
\hline\endfirsthead
 & Mode & Period(days) & ${\Delta\Omega}_{O}(mas)$ \\
\hline\endhead
\hline
\multicolumn{4}{c}{\textit{Continued on next page}}
\endfoot
\hline\endlastfoot
$\Omega_1$ & 055.565 & -404.7425 & -0.00014 \\
$S_a$      & 056.554 & 8334.8975 & 0.00051 \\
$S_{sa}$   & 057.555 & 349.8929  & 0.00013 \\
$M_m$      & 065.455 & 29.6966   & 0.00009 \\
$M_f$      & 075.555 & 14.1674   & 0.00008 \\
$M_{tm}$   & 085.455 & 9.3566    & 0.00001 \\
$2Q_1$     & 125.755 & -6.5299   & 0.00002 \\
$\sigma_1$ & 127.555 & -6.7437   & 0.00002 \\
$Q_1$      & 135.655 & -8.5579   & 0.00012 \\
$O_1$      & 145.555 & -12.4132  & 0.00071 \\
$M_1$      & 155.655 & -22.9101  & 0.00007 \\
$\pi_1$    & 162.556 & -64.2230  & 0.00005 \\
$P_1$      & 163.555 & -77.9244  & 0.00105 \\
$S_1$      & 164.555 & -99.0585  & 0.00045 \\
$K_1$      & 165.555 & -135.9224 & 0.00538 \\
$\psi_1$   & 166.554 & -216.4803 & 0.00007 \\
$\phi_1$   & 167.555 & -531.5420 & 0.00029 \\
$J_1$      & 175.455 & 34.5608   & -0.00009 \\
${Oo}_1$   & 185.555 & 15.1872   & -0.00003 \\
$2N_2$     & 235.755 & -6.3338   & -0.00029 \\
$\mu_2$    & 237.555 & -6.5349   & -0.00035 \\
$N_2$      & 245.655 & -8.2243   & -0.00197 \\
$\nu_2$    & 247.455 & -8.5665   & -0.00037 \\
$M_2$      & 255.555 & -11.7235  & -0.00817 \\
$L_2$      & 265.455 & -20.4052  & -0.00065 \\
$T_2$      & 272.556 & -49.2367  & -0.00205 \\
$S_2$      & 273.555 & -56.9078  & -0.04169 \\
$K_2$      & 275.555 & -82.6687  & -0.01707 \\
$\eta_2$   & 285.455 & 41.3306   & 0.00065 \\
\end{longtable}

\begin{longtable}{lrrr}
\multicolumn{4}{c}{{\bf Table 6}. Amplitudes $\Delta\Omega$ and periods of the perturbations for the LARES satellite}\\
\multicolumn{4}{c}{by the Moon and Sun induced tides of the ocean.}\\
\multicolumn{4}{c}{$l=4$,$p=2$,$q=0$}\\
\hline
 & Mode & Period(days) & ${\Delta\Omega}_{O}(mas)$ \\
\hline\endfirsthead
 & Mode & Period(days) & ${\Delta\Omega}_{O}(mas)$ \\
\hline\endhead
\hline
\multicolumn{4}{c}{\textit{Continued on next page}}
\endfoot
\hline\endlastfoot
$\Omega_1$ & 055.565 & 6798.3636 & 27.03953 \\
$S_a$      & 056.554 & 365.2596  & 0.25527 \\
$S_{sa}$   & 057.555 & 182.6211  & 0.80559 \\
$M_m$      & 065.455 & 27.5546   & 0.18429 \\
$M_f$      & 075.555 & 13.6608   & 0.21620 \\
$M_{tm}$   & 085.455 & 9.1329    & 0.02776 \\
$2Q_1$     & 125.755 & -6.6434   & -0.00331 \\
$\sigma_1$ & 127.555 & -6.8649   & -0.00410 \\
$Q_1$      & 135.655 & -8.7540   & -0.03139 \\
$O_1$      & 145.555 & -12.8301  & -0.23620 \\
$M_1$      & 155.655 & -24.3718  & -0.03316 \\
$\pi_1$    & 162.556 & -77.2026  & -0.03556 \\
$P_1$      & 163.555 & -97.8938  & -0.77122 \\
$S_1$      & 164.555 & -133.7391 & -0.05713 \\
$K_1$      & 165.555 & -211.0000 & -5.02186 \\
$\psi_1$   & 166.554 & -499.6108 & -0.09452 \\
$\phi_1$   & 167.555 & 1357.8063 & 0.46048 \\
$J_1$      & 175.455 & 31.6934   & 0.04318 \\
${Oo}_1$   & 185.555 & 14.6065   & 0.01162 \\
$2N_2$     & 235.755 & -6.4406   & -1.29603 \\
$\mu_2$    & 237.555 & -6.6486   & -1.57910 \\
$N_2$      & 245.655 & -8.4053   & -10.78670 \\
$\nu_2$    & 247.455 & -8.7630   & -2.08704 \\
$M_2$      & 255.555 & -12.0947  & -68.39948 \\
$L_2$      & 265.455 & -21.5567  & -2.93874 \\
$T_2$      & 272.556 & -56.5219  & -6.63498 \\
$S_2$      & 273.555 & -66.8695  & -133.12218 \\
$K_2$      & 275.555 & -105.5000 & -58.39958 \\
$\eta_2$   & 285.455 & 37.2954   & 1.03664 \\
$M_4$      & 455.555 & -6.0473   & -0.16397 \\
\end{longtable}

\section{Conclusions}

110 significant Earth's tide modes for the LARES satellite were obtained using the perturbative methods of celestial mechanics and the recent data on the satellite's orbit. The significant ocean modes were also represented. Note, that for modes of periods exceeding about a year the elasticity effects \cite{iers2010} have to be considered in more details. 
Tidal modes together with non-gravitational effects such as the Yarkovsky-Rubincam effect, are among the main orbital perturbations of laser-ranged satellites; see the randomness analysis of the residuals of the LAGEOS satellites in \cite{EPL}. 

The obtained periods and the amplitudes of the Earth's modes for LARES were represented in a way, that the mutual relevance of the modes is explicitly revealed. The detailed knowledge of the Earth's tide modes, i.e., their periods and amplitudes in the nodal longitude of LARES, LAGEOS and LAGEOS 2, is relevant to increase the accuracy in the test of frame-dragging and to estimate the uncertainty of this test and, hence, in probing General Relativity.

We acknowledge the useful comments by the referees.
I.C., A.P. and G.S. gratefully acknowledge the Italian Space Agency for the support of the LARES and LARES 2 space missions
under agreements No. 2017-23-H.0 and No. 2015-021-R.0. We are also grateful to the International Ranging Service,
ESA, AVIO and ELV.


\begin{thebibliography}{9}

\bibitem{CW} I. Ciufolini and J.A. Wheeler, {\it Gravitation and Inertia}, Princeton University Press, 1996

\bibitem{Ciu2007} I. Ciufolini, Nature {\bf 449}, 41 (2007)

\bibitem{Ciu2012} I. Ciufolini, et al., Eur. Phys. J. Plus {\bf 127}, 133 (2012)

\bibitem{Ciu2013} I. Ciufolini, V.G. Gurzadyan, R. Penrose, A. Paolozzi, in: {\it Low Dimensional Physics and Gauge
Principles}, Matinyan Festschrift, p. 93, World Scientific, 2013

\bibitem{Ciu2015} I. Ciufolini, et al., Eur. Phys. J. Plus {\bf 130}, 133 (2015)

\bibitem{Ciu2016} I. Ciufolini, et al.,  Eur. Phys. J. C, {\bf 76}, 120 (2016)

\bibitem{Br} D. Brouwer, G.M. Clemence, {\it Methods of Celestial Mechanics}, Academic Press, New York and London, 1961

\bibitem{kaula} W. M. Kaula, \textit{Theory of Satellite Geodesy}, Blaisdell Publishing Company, Waltham, 1966

\bibitem{carp89} B. Bertotti, and M. Carpino, in \textit{Measurement of the Gravitomagnetic Field
Using a Pair of Laser Ranged Satellites}, ASI final report, pp. 105, Frascati, 1989

\bibitem{Io} L. Iorio, Cel. Mech. Dyn. Astron., 79, 201 (2001)

\bibitem{iers2010} Petit, G., and Luzum, B., {\it IERS Conventions, IERS Technical Note 36}, 79 pp., Frankfurt am Main: Verlag des Bundesamts fuur
Kartographie und Geodcurrency sie, 2010

\bibitem{cata} D. E. Cartwright, R. J. Tayler, Geophys. J. R. Astron. Soc., {\bf 23}, 45 (1971)

\bibitem{caed} D. E. Cartwright, A. C. Edden, Geophys. J. R. Astron. Soc., {\bf 33}, 253 (1973)

\bibitem{love} A. E. H. Love, {\it A Treatise on the Mathematical Theory of Elasticity}, Dover, New York, 1926

\bibitem{doodson} A. T. Doodson, Proc. Roy. Soc. A., {\bf 100}, 305 (1921)

\bibitem{ijmpd} V. G. Gurzadyan, et al, Int. J. Mod. Phys. D, {\bf 26}, 1741020 (2017)

\bibitem{EPL} V. G. Gurzadyan, et al,  EPL, {\bf 102} 60002 (2013)

\bibitem{NSWC} R.D. Ray, B.V. Sanchez, \textit{Radial Deformation of the Earth by Oceanic Tidal Loading}, NASA Technical Memorandum 100743, Goddard Space Flight Center Greenbelt, Maryland 20771, (1989)

\bibitem{Cart91} D. E. Cartwright, R.D. Ray, B.V. Sanchez, \textit{Oceanic Tide Maps and Spherical Harmonic Coefficients from Geosat Altimetry}, NASA Technical Memorandum 104544, Goddard Space Flight Center Greenbelt, Maryland 20771, (1991)

\bibitem{Ray1999} R. D. Ray, \textit{A Global Ocean Tide Model From TOPEX/POSEIDON Altimetry: GOT99.2}, NASA/TM-1999-209478, Goddard Space Flight Center Greenbelt, Maryland 20771, (1999)

\bibitem{Wu2012} B. Wu, \textit{Comparison of different ocean tide models especially with respect to the GRACE satellite mission}, Studienarbeit im Studiengang Geod{\"a}sie und Geoinformatik an der Universit{\"a}t Stuttgart, Universit{\"a}t Stuttgart Geod{\"a}tisches Institut, (2012)

\bibitem{GEMT1} C. Christodoulidis, D. E. Smith, R. G. Williamson, S. M. Klosko, \textit{Observed Tidal Earth/Moon/Sun Braking in the System}, NASA Technical Memorandum 100677, Goddard Space Flight Center Greenbelt, Maryland 20771, (1987)

\bibitem{EGM96} F.G. Lemoine, et al., \textit{The Development of the Joint NASA GSFC and the National Imagery and Mapping Agency (NIMA) Geopotential Model EGM96}, NASA/1998, 206861, Goddard Space Flight Center Greenbelt, Maryland 20771, (1998)  


\bibitem{Farrel72} 	W. E. Farrell, Revs. Geophys. Space Phys., {\bf 10}, 761 (1972)

\bibitem{NA11} S.-H. Na, J. Baek, \textit{Computation of the Load Love Number and the Load Green's Function for an Elastic and Spherically Symmetric Earth}, J. Korean Phys. Soc., {\bf 58} 5 (2011)

\bibitem{FES2004} F. Lyard, F. Lefevre, T. Letellier, O. Francis, Ocean Dynamics {\bf 56}, 394 (2006)





  
\end{thebibliography}
\end{document}